# 3-D Reconfigurable Intelligent Surface: From Reflection to Transmission and From Single Hemisphere to Full 3-D Coverage

Ruiqi Wang, Yiming Yang, *Student Member, IEEE*, and Atif Shamim, *Fellow, IEEE*

*Abstract*—Reconfigurable intelligent surfaces (RIS) are conventionally implemented as two-dimensional (2D) electromagnetic (EM) structures to steer incident waves toward desired reflection angles. This approach limits the reflection to a single hemisphere and the beam-scanning range is rather small. In this work, a novel three-dimensional (3D) RIS concept is proposed, where beam-scanning can be realized not only through reflection from the illuminated surface but also through controlled transmission toward adjacent surfaces, enabling near blind-spot-free coverage in the full 3D spatial domain. A cube-based 3D-RIS design, operating at millimeter-wave (mm-Wave) frequencies and consisting of six interconnected RIS is presented in this work. Each surface integrates reconfigurable receiving and reflection arrays with orthogonal polarizations to ensure intrinsic EM isolation, while a reconfigurable feeding network is employed to support dynamic operation. A subarray-based synthesis approach with binary amplitude gating and predefined phase offsets is developed through a unified theoretical model. This theoretical model, validated through full-wave simulations, enables efficient beam switching through a shared aperture. Based on this framework, an 8 × 12-element surface comprising six 4 × 4 subarrays is designed, with each surface covering an angular range from -30° to +30°. The experimental prototype has been characterized in the 24 – 30 GHz band and the measured results demonstrate a gain enhancement of 14.7 dB for reflection, while 14.1 dB has been achieved for transmission to the neighboring-surface. Finally, wireless communication trials using the Pluto software-defined radio platform combined with frequency up/down converters confirm improved constellation quality and a 6–7 dB improvement in error vector magnitude (EVM) for both reflection and neighboring-surface transmission scenarios.

*Index Terms*—Millimeter-wave (mm-Wave), Reconfigurable intelligent surface (RIS), three-dimensional (3D), sixth-generation (6G), subarray synthesis, wideband, wireless communication.

## I. INTRODUCTION

RECONFIGURABLE intelligent surfaces (RIS) have emerged as a key enabling technology for sixth-generation (6G) and beyond wireless communication systems due to their ability to reconfigure the electromagnetic (EM) environment in a programmable and energy-efficient manner [1]. By intelligently manipulating the phase, amplitude, or polarization of incident EM waves, RIS can reshape wireless propagation channels, enhance signal coverage, and improve link reliability [2]. One of the most important advantages of RIS lies in its capability to establish secondary line-of-sight (LoS) links when the direct path between the transmitter and receiver is blocked, thereby mitigating the detrimental effects of shadowing and blockage in dense wireless environments [3]. Moreover, RIS has been widely investigated in conjunction with advanced wireless technologies such as multiple-input multiple-output (MIMO) systems [4], fluid antenna systems (FAS) [5], and integrated sensing and communication (ISAC) frameworks [6], highlighting its broad applicability in future intelligent wireless networks.

Despite the rapid growth of RIS-related research, a large portion of existing studies remains primarily theoretical and relies on idealized assumptions that are difficult to realize in practice [7–9]. From a hardware perspective, most experimentally demonstrated RIS prototypes have been reported at below the mm-Wave regime [10–22]. Looking ahead, wireless systems are expected to increasingly leverage higher-frequency bands to access larger bandwidths and support higher data rates. In this regard, mm-Wave and sub-terahertz frequencies are envisioned as key enablers for 5G-Advanced and future 6G systems [23], where wireless propagation becomes increasingly vulnerable to blockage and non-line-of-sight (NLoS) conditions, making the deployment of practical mm-Wave RIS particularly critical. Nevertheless, mm-Wave RIS hardware design remains challenging due to stringent requirements on bandwidth, insertion loss, phase quantization accuracy, and switching component performance. In recent years, several efforts have been devoted to advancing mm-Wave RIS hardware. Representative works have investigated wideband RIS designs [24], mitigation of sidelobe and quantization-lobe issues [25], multimode resonances [26], and performance improvement of PIN-diode-based switching networks at mm-Wave frequencies [27], [28]. Other studies have explored independent amplitude and phase control [29], as well as fully printed, low-cost RIS prototypes [30]. While these works have significantly advanced the state of the art in mm-Wave RIS design, they are predominantly based on conventional planar two-dimensional (2D) surface configurations. Although simultaneously transmitting and reflecting RIS (STAR-RIS), have been proposed to enable signal manipulation on both sides of a planar surface [31]. However, practical STAR-RIS hardware still suffers from limited spatial coverage and unavoidable blind regions, especially near endfire directions [32].

In order to alleviate the above mentioned limitations, this work proposes a 3D-RIS that extends EM wave manipulation from a single surface to the volumetric domain. By enabling not only reflection on the illuminated surface but also controlled signal transfer toward adjacent faces, the proposed architecture effectively mitigates blind regions that persist in conventional 2D and STAR-RIS designs. To maintain scalability and practical implementability, each RIS face adopts a subarray-based architecture with binary amplitude gating and predefined phase offsets, which reduces control complexity while retaining flexible beam synthesis capability. Based on these principles, a unified theoretical framework is established to describe subarray-level beamforming and inter-surface signal routing, which directly guides the hardware realization and experimental evaluation. The proposed 3D-RIS is further validated through comprehensive measurements and over-the-air wireless communication experiments, demonstrating its practical feasibility and system-level relevance. In summary, the main contributions of this work are as follows:

1) A novel 3D-RIS architecture is proposed, enabling volumetric EM wave manipulation through both surface reflection and inter-surface transmission, thereby overcoming the spatial coverage limitations of conventional planar RIS designs.

2) A theoretical framework for 3D-RIS modeling is developed, incorporating subarray-based beamforming, binary amplitude gating, predefined phase offsets, and inter-surface signal routing.

3) A practical mm-Wave 3D-RIS hardware is designed, simulated and fabricated.

4) Two-horn measurements and over-the-air communication trials are conducted to validate the proposed concept from electromagnetic and communication-system perspectives.

The remainder of this paper is organized as follows. Section II presents the theoretical framework of the proposed 3D-RIS. Section III develops the ideal RIS beamswitching modeling and theory validation. Section IV describes the practical 3D-RIS design and implementation. Section V reports fabrication and experimental validation, including measurements and over-the-air communication trials. Finally, Section VI concludes the paper.

## II. THEORY

The proposed 3D-RIS concept is shown in Fig. 1, which is implemented as a cube-shaped structure consisting of six planar RIS surfaces, corresponding to the six faces of the cube. Different from traditional RIS designs that rely on element-wise reflection phase control to manipulate the reflected wavefront, the proposed architecture organizes RIS elements into multiple subarrays that collectively enable subarray-level wave manipulation across different spatial directions. The electromagnetic response of the overall RIS is reconfigured through switch-controlled interconnections among RIS elements and subarrays. To enable rapid subarray-level synthesis compared with computationally intensive full-wave simulations, and to provide a tractable interface for communication-oriented system modeling and performance evaluation, the detailed theoretical 3D-RIS framework is demonstrated as follows.

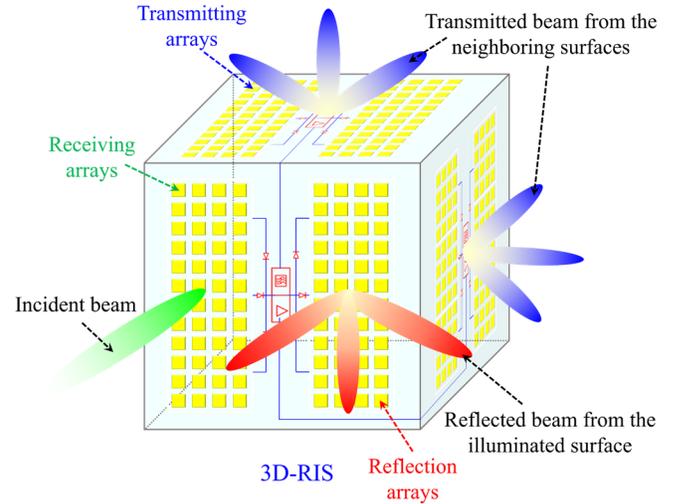

Fig. 1. The proposed 3D-RIS concept.

Let $s \in \{1, 2,\ldots, S\}$ donate the face index. Each face $s$ is associated with a local coordinate system $(x_s, y_s, z_s)$ where $x_s$ is the outward normal direction of that face, and $(y_s, z_s)$ span the face plane. The global coordinate system is denoted by $(x, y, z)$. The orientation of face $s$ is described by a rotation matrix $\mathbf{R}_s$ that maps a local vector into the global frame, namely $\mathbf{v} = \mathbf{R}_s \mathbf{v}_s$. Accordingly, the global position of an element on face $s$ follows from its local position through the same mapping.

For each surface of the 3-D RIS, it starts from a physically interpretable microstrip patch radiation model to enable analytical array synthesis and overall radiation optimization. The radiating element is modeled using the cavity model of microstrip patch formulation, where the dominant radiating mechanism is represented by an equivalent magnetic current on the patch edges. The arrangement of the RIS element in Cartesian coordinate system is demonstrated in Fig. 2. Therefore, the electric field for the microstrip patch antenna in the local coordinate of face $s$ is written as [33]:

$$E_{elem}(\theta_s,\varphi_s) = E_0 \left\{\sin\theta_s \frac{\sin(X)}{X}\frac{\sin(Z)}{Z}\right\} \times \cos(\frac{k_0 L_e}{2}\sin\theta_s \sin\phi_s) \quad (1)$$

where

$$X = \frac{k_0 h}{2}\sin\theta_s \cos\phi_s \quad (2)$$

$$Z = \frac{k_0 W}{2}\cos\theta_s \quad (3)$$

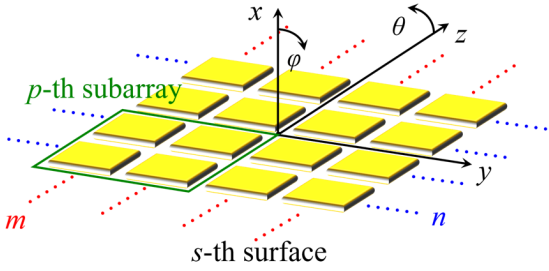

Fig. 2. The RIS element configuration.

$E_0$ denotes the field amplitude normalization constant $k_0 = 2\pi/\lambda$, $\lambda$ is the free space wavelength, $h$ denotes the substrate thickness, $W$ denotes the effective radiating aperture width, $L_e$ is the distance between the equivalent two magnetic currents. The angles $(\theta_s, \phi_s)$ are defined in the local coordinate system of face $s$. To relate the global observation direction $\hat{\mathbf{u}}(\theta, \varphi)$ to the local angles $(\theta_s, \phi_s)$, the direction unit vector is first expressed in the local frame as $\hat{\mathbf{u}}_s = \mathbf{R}_s^T \hat{\mathbf{u}}$. Since each face is associated with a local coordinate system whose $x_s$ axis represents the outward normal direction, only observation directions satisfying a positive projection onto $x_s$ are considered. Observation directions with negative projection onto the local normal direction are excluded, which corresponding to $\hat{\mathbf{u}}_s \cdot \hat{\mathbf{x}}_s < 0$.

Then, the RIS array design can be constructed. To reduce control complexity while maintaining beamforming capability, each RIS surface is partitioned into multiple subarrays. Within a given face $s$, let $p \in \{1, 2,…,P_s\}$ denote the subarray index, where each subarray consists of $N \times N$ RIS elements and serves as a beamforming unit. Let $\mathbf{r}_{s,p,0}$ be the global position of the subarray center of $(s, p)$. The local in plane displacement is $\mathbf{d}_{n,m} = [0, nd_1, md_2]^T$ in the $(x_s, y_s, z_s)$ frame. Then the global position of the $(n, m)$ element is

$$\mathbf{r}_{s.p,n,m} = \mathbf{r}_{s.p,0} + \mathbf{R}_s \mathbf{d}_{n,m} \quad (4)$$

where $d_1$ and $d_2$ denote the element spacings along the two tangential axes of the face plane, and $(n,m)$ are referenced to the subarray center. Within each subarray, a predefined progressive phase distribution is imposed to steer the reflected wave toward a desired direction $(\theta_{s,p}, \varphi_{s,p})$. For the $(n, m)$-th element inside subarray $(s, p)$, the assigned phase is given by

$$\psi_{s,p}(n,m) = -k_0(nd_1 \sin\theta_{s,p}\cos\phi_{s,p} + md_2\sin\theta_{s,p}\sin\phi_{s,p}) \quad (5)$$

In the proposed architecture, each subarray is treated as a beamforming primitive controlled by a single binary amplitude gate, which corresponds to the switching states of practical PIN diodes used to regulate the electromagnetic reflection behavior of the surface. The conduction and isolation states of the PIN diodes naturally correspond to binary amplitude states. Accordingly, a binary amplitude gate $A_{s,p} \in \{0,1\}$, is assigned to the $(s, p)$ subarray surface, where $A_{s,p} = 1$ represents the conducting state and $A_{s,p} = 0$ represents the isolated state of the PIN diode.

Moreover, in addition to binary amplitude gating, a predefined global phase offset $\Phi_{s,p}$ is introduced for each subarray surface, which represents a fixed phase bias implemented through passive structural or routing asymmetries in the hardware design. The primary motivation for incorporating subarray-level phase offsets is to enable multimode beam synthesis when multiple subarray surfaces are simultaneously activated. By properly selecting the phase offsets $\Phi_{s,p}$, different reflected beams can be coherently synthesized through the superposition of multiple subarrays. This design strategy offers important advantages. First, multiple beam patterns can be generated using a limited number of switching states, which reduces the required number of PIN diodes and control channels. More importantly, the aperture is shared among different beamforming configurations, enabling efficient utilization of the RIS hardware and supporting multiple operational modes within the same physical structure. The total excitation coefficient of the $(n,m)$-th element on the $s$-th surface and $p$-th subarray is therefore expressed as

$$w_{s,p,n,m} = A_{s,p} e^{j\Phi_{s,p}} e^{j\psi_{s,p}(n,m)} \quad (6)$$

When the incident surface is illuminated, the EM wave can either be reflected by the incident surface or transferred to adjacent cube faces through switch-controlled inter-surface connections. These switches determine whether the electromagnetic energy is allowed to propagate toward an adjacent surface. When the switch is enabled, the incident EM wave is allowed to propagate toward the adjacent surface, which follows the same element radiation model, subarray partitioning, and control scheme as the incident surface. When the switch is disabled, the adjacent surface remains electromagnetically isolated. This mechanism enables dynamic control of wave propagation across different cube faces while preserving a unified RIS design across all surfaces.

Inter surface signal transfer is governed by switch controlled paths at cube edges. Define a binary switch state $b_{s \to t} \in \{0,1\}$ for each ordered pair of adjacent faces $(s,t)$. When $b_{s \to t} = 1$, the EM wave is allowed to propagate from face $s$ toward the adjacent face $t$ through the designed transmission path. When $b_{s \to t} = 0$, the propagation toward that adjacent face is blocked, and the adjacent face stays electromagnetically isolated from that transfer path. Under a given illumination, these switch states determine a participating face set that contributes to the overall reradiated field. Therefore, the overall array factor of the proposed 3-D RIS is obtained by summing the contributions from all active elements across all subarrays

$$AF(\theta,\varphi) = \sum_s \sum_{p=1}^{P_s} \sum_{n=1}^{N} \sum_{m=1}^{N} w_{s,p,n,m} \times e^{jk_0 \hat{\mathbf{u}}^T \mathbf{r}_{s,p,n,m}} \quad (7)$$

The far field of the cube based 3D RIS is then expressed as a coherent superposition of the rotated element patterns weighted by the global array factor.

$$E(\theta,\varphi) = \sum_s E_{elem}(\theta_s,\varphi_s) \sum_{p=1}^{P_s} \sum_{n=1}^{N} \sum_{m=1}^{N} w_{s,p,n,m} \times e^{jk_0 \hat{\mathbf{u}}^T \mathbf{r}_{s,p,n,m}} \quad (8)$$

It is worth noting that the proposed RIS surface is physically configured to operate with two distinct functional regions, namely the receiving region and the reflection region. These two regions are implemented using orthogonal polarizations, which provides intrinsic EM isolation between the receiving and reflection processes on the same physical surface. As a result, the incident wave captured by the receiving region does not directly interfere with the reradiated field produced by the reflection region, thereby enabling independent and stable operation of the two functionalities. In this work, the receiving region serves to receive the incident EM wave into the RIS structure and route the signal toward the reflection region through the designed interconnections, while the reflection region is responsible for shaping the outgoing wavefront according to the prescribed subarray-based beamforming strategy. Therefore, the overall RIS receiving and reflection pattern formulation can be expressed as

$$E_{rec}(\theta,\varphi) = E(\theta,\varphi); E_{ref}(\theta,\varphi) = E_{\perp}(\theta,\varphi) \quad (9)$$

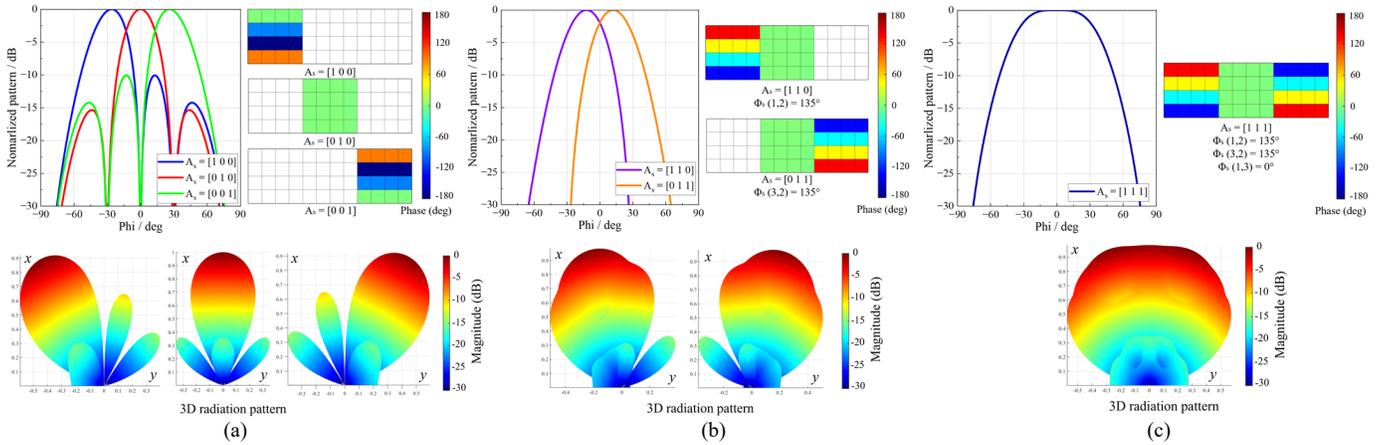

Fig. 3. Theoretical calculation of the 2D and 3D normalized radiation patterns and array phase distributions of the RIS beamswitching model. (a) Beamforming configuration 1. (b) Beamforming configuration 2. (c) Beamforming configuration 3.

Consequently, the radiation patterns for both receiving and reflection regions on the 3-D RIS follow the same analytical formulation given in (8), with the only difference being the polarization state of the radiated field. Specifically, the reflected fields are polarized orthogonally with respect to the receiving polarization, while preserving identical element radiation characteristics, subarray partitioning, excitation coefficients, and array factor expressions. This polarization-orthogonal design ensures consistent beamforming behavior across all reflecting surfaces while maintaining effective isolation between receiving and reflection operations.

## III. IDEAL RIS DESIGN AND VALIDATION OF THEORY

In this section, design considerations are discussed and an idealized RIS configuration is introduced to validate the theoretical framework developed in Section II. Practical non-idealities such as the feeding network and switches are excluded, enabling direct analytical evaluation and clear comparison with full-wave simulations, thereby providing transparent verification of the proposed theory before moving to practical hardware implementation.

### A. Design Considerations

Based on the general theoretical framework proposed in Section II, this subsection presents a practical proof-of-concept configuration where one cube face is considered as the illuminated surface, and the analytical synthesis is carried out in closed form using the subarray-based weighting strategy for the 3D cube RIS model in (5)–(8).

In the designed model, the mm-Wave frequency band with a center frequency of $f_0$ = 26 GHz is selected considering future 5G/6G integrated system demands for high operating frequency with naturally large raw bandwidth. The distance between two array elements is selected as half-wavelength to avoid the generation of grating lobes at various beamforming angles. For the subarray construction, each subarray consists of $N \times N$ elements, where $N = 4$ is adopted in this configuration for proof-of-concept purposes, considering the tradeoff among beamforming gain, radiation beamwidth, and feeding network complexity. Smaller subarrays would suffer from limited directivity and excessive beam spreading, whereas significantly larger subarrays would introduce overly narrow beams and reduce angular overlap between adjacent beamforming configurations.

In the proof-of-concept configuration, three subarrays are stacked along the z-axis of the illuminated surface, and each subarray is assigned a predefined steering direction in the azimuth plane (xy-plane). An angular separation of 30° is determined in this configuration, which ensures sufficient spatial separation between adjacent beams, allowing the individual beamforming behaviors of different subarrays to be clearly distinguished. Meanwhile, it provides decent radiation coverage under the 4 × 4 subarray structure. Therefore, the overall array is constructed with three 4×4 subarrays with beam-switching angles of $\phi$ = −30°, 0°, and 30°, while the elevation angle is fixed at $\theta$ = 90°. It should be noted that this work is designed with 1-D beam scanning. However, the general formulation in Section II naturally supports subarray arrangements with 2-D beam-scanning capability on each RIS surface. Therefore, when each subarray is activated individually, the amplitude gate matrix $A_s$ can be expressed as [1, 0, 0], [0, 1, 0], and [0, 0, 1], where the calculated theoretical 2-D and 3-D normalized radiation patterns and array phase distributions are demonstrated in Fig. 3(a). From the results, it can be observed that when each subarray is activated, the beamforming angles have three distinct directions around −30°, 0°, and 30°, which serve as the fundamental beamforming configuration of the RIS design. These fundamental operation configuration construct the basic RIS beam-scanning capability.

Although the beam-scanning resolution can be increased as the number of discrete subarray states increases, this also increases the fabrication cost, aperture area, and control complexity. To achieve an efficient RIS design without adding additional design complexity, the shared-aperture concept is adopted in this work, where the phase offset $\Phi_{s,p}$ is loaded to each subarray. Such an arrangement aims to achieve effective overall RIS array synthesis. In this design, it is optimized that the same magnitude and a 135° phase difference between two adjacent subarrays can realize approximately ±15° beam-scanning capability. In other words, when the subarray amplitude gate matrix $A_s$ = [1, 1, 0], and the phase offset difference between the +30° beamforming subarray and the 0° beamforming subarray satisfies $\Phi_s(1,2)$ = 135°, an intermediate beamforming angle of +15° can be obtained. The theoretical normalized 3-D and 2-D radiation patterns are illustrated in Fig. 3(b). Similarly, a −15° beamforming angle can be realized by simultaneously activating the other two subarrays with amplitude gate matrix $A_s$ = [0, 1, 1] and phase $\Phi_s(3,2)$ = 135°. The physical mechanism behind the beam pattern synthesis is proper electric-field phase matching between adjacent subarrays. Therefore, the phase offset term $\Phi_{s,p}$ enables the desired phase alignment. Hence, when all subarrays are activated with amplitude gate matrix $A_s$ = [1, 1, 1] and phase conditions $\Phi_s(1,2)$ = 135°, $\Phi_s(3,2)$ = 135°, and $\Phi_s(1,3)$ = 0°, a wide-beamwidth radiation pattern can be achieved, as demonstrated in Fig. 3(c). Under this working configuration, the RIS provides wide-angle beam coverage for signals arriving from various base stations or users.

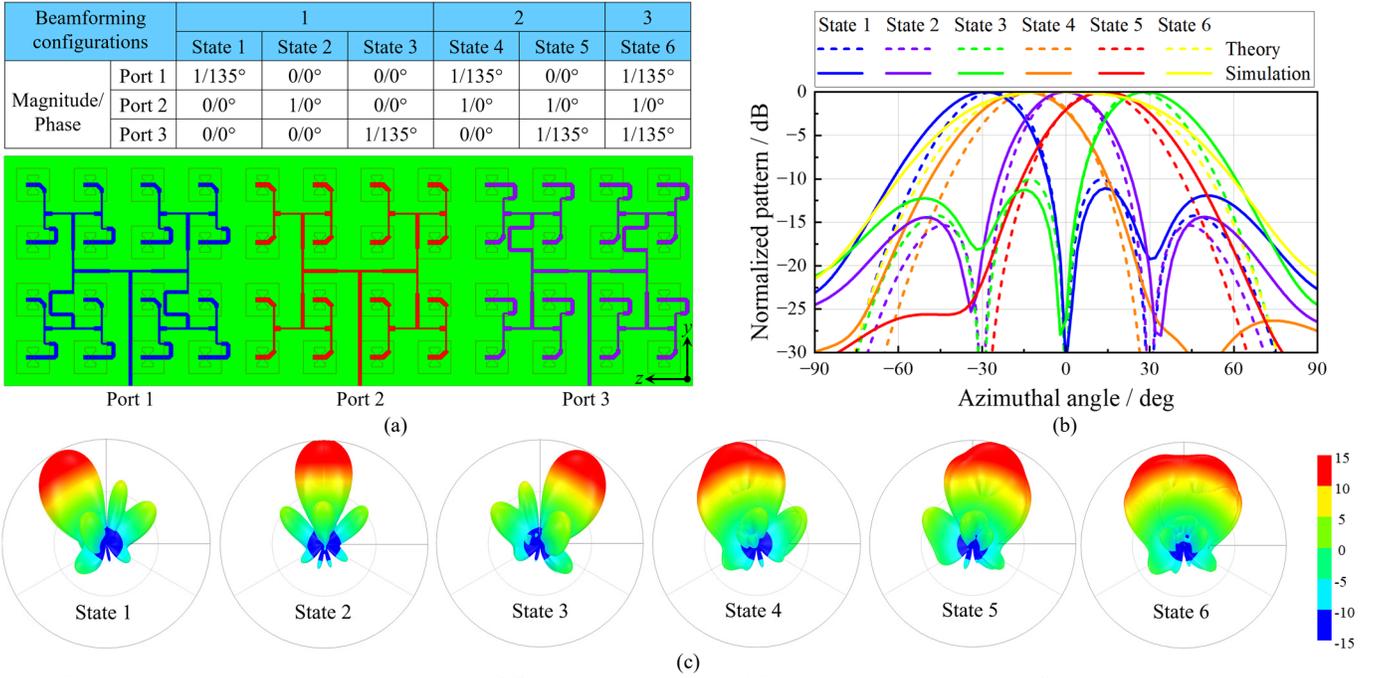

Fig. 4. Full-wave simulation validation for the theoretical RIS beamswitching model. (a) Full-wave simulation model; (b) Comparison between the calculated and simulated normalized 2D radiation pattern; (c) Simulated 3D radiation pattern.

Therefore, such a practical RIS array model can be utilized as a beam-scanning array containing three working configurations. The fundamental configuration enables independent array beamforming toward different directions through discrete subarray units. The second configuration activates multiple subarrays to generate additional beamforming angles, which compensate for the beam coverage of the fundamental configuration and thus increase the beam pattern resolution through the shared-aperture concept. Finally, the third configuration achieves wide-beamwidth coverage by activating all subarrays. For each single surface of the RIS design, two such beam-scanning arrays with orthogonal polarization arrangements are employed. One RIS array is used for receiving, while the orthogonally polarized array is used for reflection, as described in (9). Hence, a single surface of the 3D-RIS can be constructed. The complete 3D-RIS structure is then formed by interconnecting multiple such single RIS surfaces to achieve spatial beam coverage. It should be pointed out that this proof-of-concept configuration demonstrates an efficient 3D-RIS design, and more advanced 3D RIS architectures can be explored based on the proposed theoretical framework.

B. *Full-Wave Simulations and Comparison with Theory*

To validate the proposed theoretical RIS beam switching model developed in Section III-A, full-wave EM simulations are carried out using ANSYS HFSS. The simulation setup is aligned with the analytical configuration in Fig. 3, such that a direct comparison between the theoretical predictions and the full-wave results can be established.

The full-wave simulation model is illustrated in Fig. 4(a). The simulated structure follows the same subarray partitioning strategy adopted in the theoretical analysis, where the RIS surface is divided into three identical 4×4 subarrays stacked along the z-direction. Each subarray is assigned a predefined progressive phase distribution corresponding to steering angles of $\phi$ = −30°, 0°, and +30° at a fixed elevation angle of $\theta$ = 90°. The operating frequency is set to $f_0$ = 26 GHz, and the inter-element spacing is maintained at half-wavelength to suppress

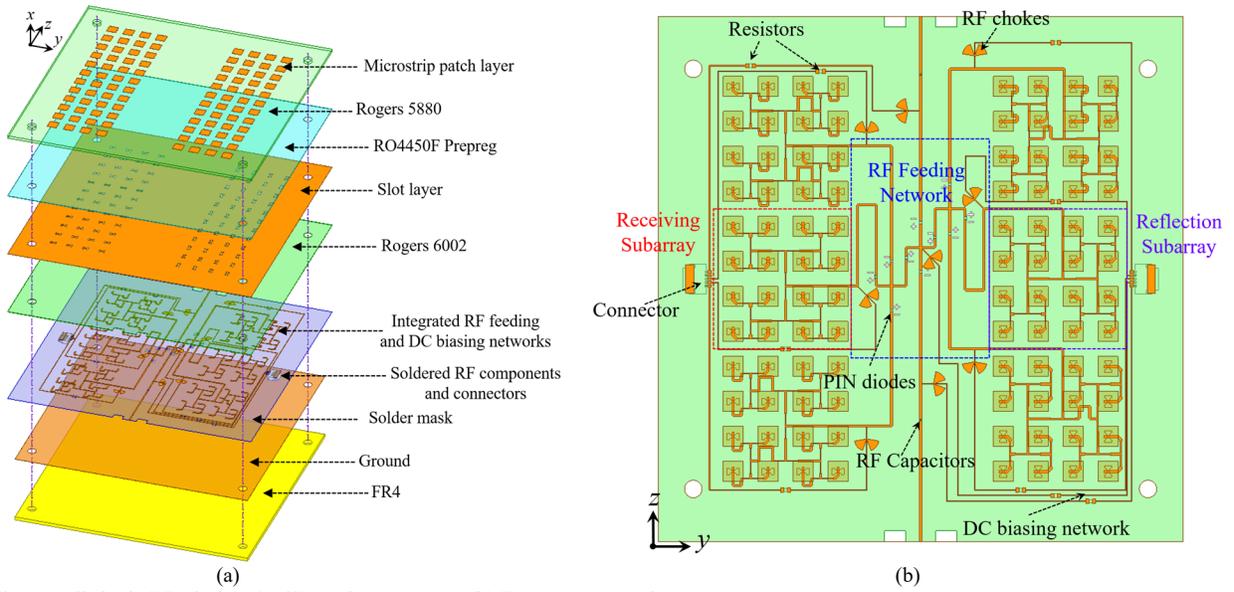

Fig. 5. The overall single-RIS design. (a) 3D stackup structure. (b) Transparent top view.

grating lobes, consistent with the assumptions in the analytical model.

In the full-wave simulation, the subarray-level amplitude gates and phase offsets are implemented through discrete excitation ports, emulating the effective excitation coefficients defined in (6). Different beam-switching configurations are realized by activating specific combinations of subarrays and imposing the corresponding relative phase offsets, matching the theoretical configurations 1, 2, and 3 presented in Fig. 3. Fig. 4(b) compares the normalized 2D radiation patterns obtained from the theoretical calculations and the full-wave simulations for all beam-scanning states. A close agreement is observed in terms of main-beam direction and relative sidelobe levels. Minor discrepancies are mainly attributed to mutual coupling effects in the practical layout, which are captured in the full-wave simulation. Furthermore, the 3D simulated radiation patterns for different beamswitching states are presented in Fig. 4(c). The results confirm that the proposed subarray-based beam-switching mechanism is capable of generating distinct directional beams as well as wide-beam coverage configuration through shared-aperture excitation. The overall spatial radiation characteristics observed in the full-wave simulations are consistent with the theoretical predictions in Fig. 3, thereby validating the accuracy and physical relevance of the proposed RIS beam-switching model.

## IV. PRACTICAL 3D RIS DESIGN

### A. Single surface Design and Simulations

Based on the analytical 3-D RIS framework developed in Section II and Section III, this subsection presents a practical hardware realization of a single-surface RIS, which serves as the fundamental building block of the proposed 3-D RIS architecture. Each surface is independently functional, enabling seamless extension toward a cube-based 3D configuration.

The overall architecture of the single-surface RIS is illustrated in Fig. 5. As shown in Fig. 5(a), the proposed surface adopts a multilayer printed circuit board stackup that integrates radiating elements, RF signal routing, and DC biasing networks within a compact footprint. The radiating layer consists of microstrip patch elements implemented on a Rogers 5880 substrate, while an aperture-coupled slot layer based on Rogers 6002 is employed to facilitate EM coupling between the receiving and reflection regions. The RO4450F prepreg layer is used to sandwich the PCB layers and ensure mechanical stability. The bottom FR4 layer provides the ground reference as well as mechanical support for the entire RIS structure. A transparent top view of the single-surface RIS is shown in Fig. 5(b), where the subarray-based architecture is clearly visible. Consistent with the theoretical formulation in Section III, the surface is partitioned into three identical subarrays arranged along one spatial dimension. Each subarray functions as a beamforming primitive and is controlled by a single binary amplitude gate implemented using PIN diode. In this work, the MADP-000907-14020P PIN diode is adopted due to its favorable RF performance in the millimeter-wave frequency band and its relatively low cost. This subarray-level control strategy significantly reduces the number of switching components and control lines, while preserving the beam synthesis capability predicted by the analytical model.

The detailed geometrical configuration of the single-surface RIS is illustrated in Fig. 6. Fig. 6(a) shows the unit-cell structure, which is based on an aperture-coupled microstrip patch configuration to enable stable radiation characteristics and polarization manipulation. Fig. 6(b) depicts the subarray layout composed of multiple unit cells, which directly corresponds to the beamforming primitives defined in Section III. Fig. 6(c) presents the RF feeding network, where the routing topology and passive phase paths are designed to implement the predefined subarray-level phase offsets. This feeding architecture enables the realization of different beamforming configurations through shared-aperture excitation without introducing additional phase shifters. All critical geometrical parameters of the single-surface RIS are summarized in Table

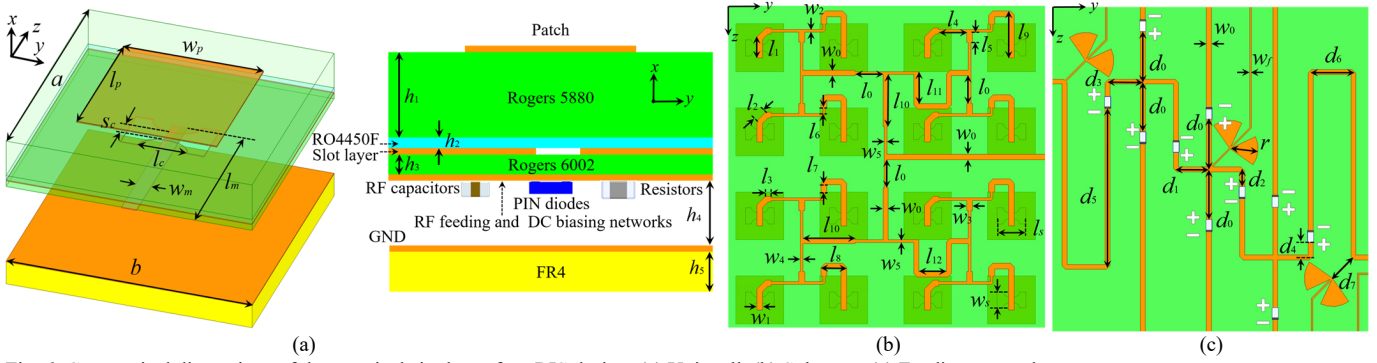

Fig. 6. Geometrical dimensions of the practical single-surface RIS design. (a) Unit cell. (b) Subarray. (c) Feeding network.

TABLE I. DIMENSIONS OF THE SINGLE SURFACE OF THE 3D-RIS (mm)

| $a$ | $b$ | $l_p$ | $w_p$ | $l_m$ | $l_c$ | $s_c$ | $h_1$ | $h_2$ | $h_3$ | $h_4$ | $h_5$ | $l_s$ | $w_s$ | $w_m$ | $l_0$ | $l_1$ | $l_2$ | $l_3$ | $l_4$ | $l_5$ | $l_6$ |
|---|---|---|---|---|---|---|---|---|---|---|---|---|---|---|---|---|---|---|---|---|---|
| 5.0 | 5.0 | 2.85 | 2.85 | 3.1 | 0.99 | 0.295 | 0.787 | 0.101 | 0.127 | 1.6 | 0.5 | 1.7 | 0.955 | 0.35 | 1.67 | 1.25 | 0.742 | 0.326 | 1.65 | 0.596 | 0.337 |
| $l_7$ | $l_8$ | $l_9$ | $l_{10}$ | $l_{11}$ | $l_{12}$ | $w_0$ | $w_1$ | $w_2$ | $w_3$ | $w_4$ | $w_5$ | $w_f$ | $d_0$ | $d_1$ | $d_2$ | $d_3$ | $d_4$ | $d_5$ | $d_6$ | $d_7$ | $r$ |
| 0.422 | 1.526 | 2.792 | 3.1 | 1.906 | 1.638 | 0.31 | 0.35 | 0.16 | 0.35 | 0.16 | 0.163 | 0.1 | 3.35 | 2.07 | 1.28 | 2.21 | 1.14 | 10.13 | 2.69 | 1.87 | 1.77 |

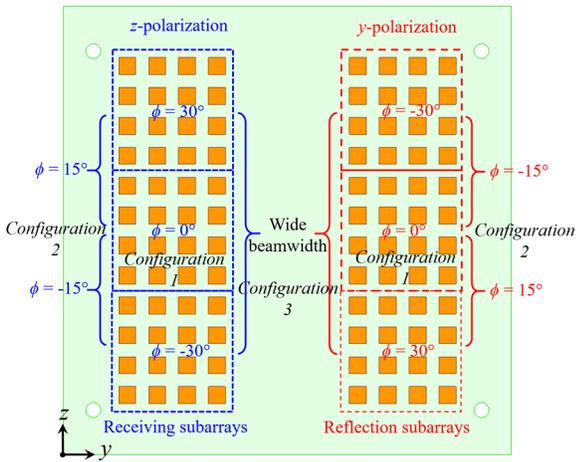

Fig. 7. Working principle of the practical single-surface RIS design.

I. It should be noted that the listed subarray dimensions correspond to a representative design optimized for the receiving-mode beamforming toward +30°. The remaining subarrays are optimized following the same design methodology, with slight dimensional variations to realize their respective beamforming functionalities. For brevity, the detailed dimensions of these subarrays are not explicitly listed.

The working principle of the single-surface RIS is conceptually illustrated in Fig. 7. Upon illumination, the incident EM wave is first captured by the receiving subarray and guided through the designed RF interconnections toward the reflection subarray. The binary states of the PIN diodes determine whether each subarray is activated or isolated, thereby implementing the amplitude gating coefficients defined in Section III. When a subarray is activated, its predefined progressive phase distribution shapes the reflected wavefront toward the desired direction. By selectively enabling different subarrays or combinations thereof, the single-surface RIS supports multiple beamforming configurations, including discrete beam steering (Configuration 1), intermediate beam synthesis through shared-aperture excitation (Configuration 2), and wide-beam coverage (Configuration 3).

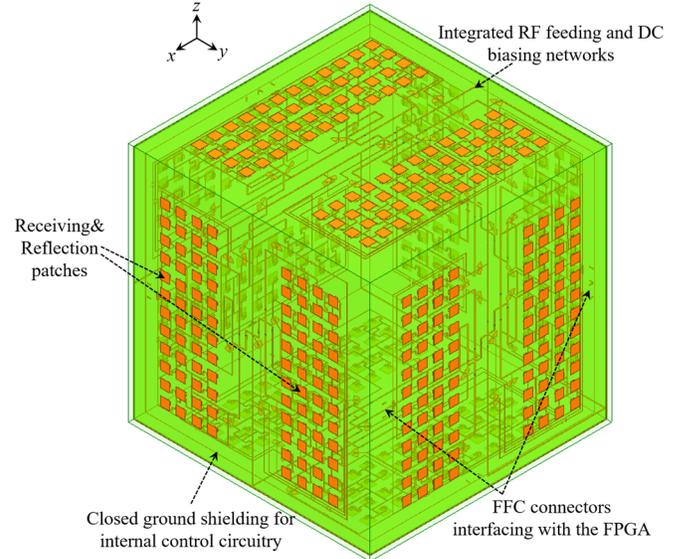

Fig. 8. The proposed 3D-RIS design.

### B. Full 3D RIS Simulations

Based on the single-surface RIS module described in Section IV-A, the proposed 3D RIS is constructed by integrating multiple RIS surfaces into a cubic configuration, as illustrated in Fig. 8. Each face of the cube corresponds to an independently functional RIS surface, while the overall structure enables volumetric wave manipulation through both surface reflection and inter-surface transmission. This modular architecture directly follows the analytical framework developed in Section II and Section III, where each RIS face is associated with its own local coordinate system and beamforming primitives, while the global radiation behavior emerges from the coherent interaction among multiple faces.

Fig. 8 shows the complete 3D-RIS structure, where six single-surface RIS are arranged to form a closed cube. The integrated RF feeding and DC biasing networks are integrated within each piece of RIS surface (Fig. 5), and each ground plane of the RIS forms a closed shielding space which can accommodate internal control circuitry to mitigate undesired EM interference. Flexible flat cable (FFC) connectors are employed to interface the RIS control network with an external FPGA controller, enabling independent and reconfigurable control of each surface and subarray. This hardware configuration preserves the scalability of the proposed design, allowing additional surfaces or alternative geometries to be realized using the same single-surface building block.

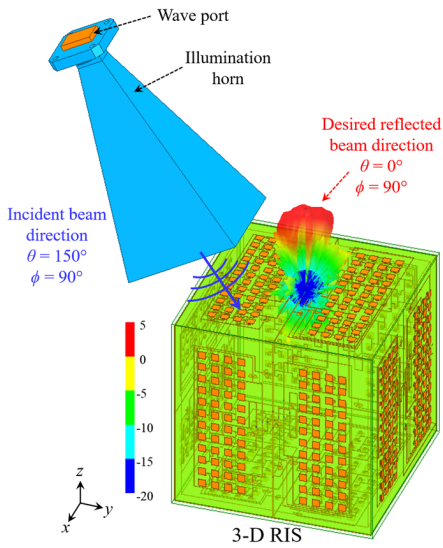

Fig. 9. The proposed 3D-RIS design with single-surface reflection.

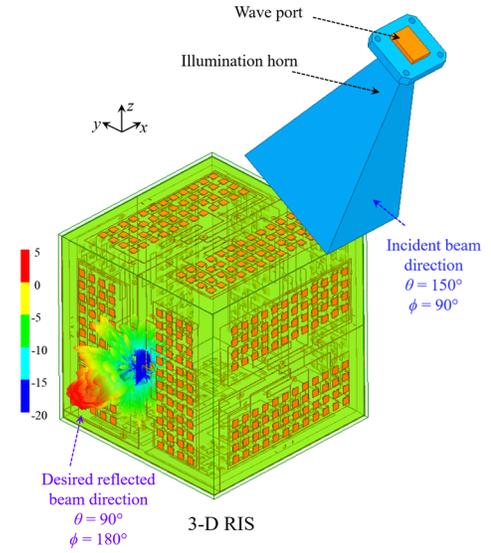

Fig. 11. The proposed 3D-RIS design with neighboring-surface transmission.

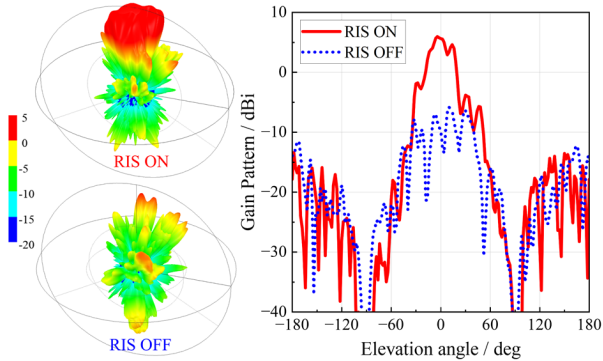

Fig. 10. Simulated 3D and 2D gain patterns for the proposed 3D-RIS with single-surface reflection in the ON and OFF states.

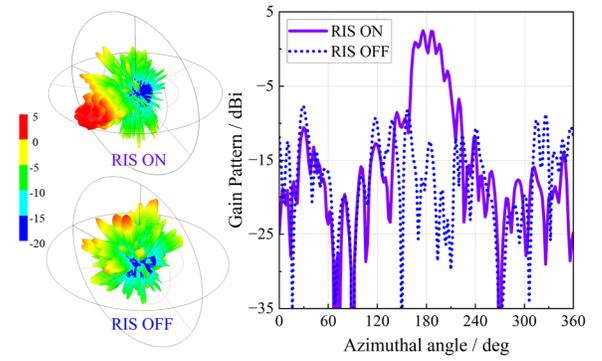

Fig. 12. Simulated 3D and 2D gain patterns for the proposed 3D-RIS with neighboring-surface transmission in the ON and OFF states.

To first demonstrate the fundamental operation of the 3D-RIS, a single-surface reflection scenario is considered, as illustrated in Fig. 9. In this configuration, the incident EM wave illuminates one face of the cube, while only the corresponding reflection subarrays on that surface are activated. All other surfaces remain EM isolated. The desired reflected beam direction is synthesized by activating the appropriate subarrays on the illuminated face, following the same subarray-level amplitude gating and phase distribution strategy described in Section III.

The simulated 3D and 2D gain patterns for the single-surface reflection case are presented in Fig. 10. When the RIS is in the ON state, a clear directive beam is formed along the designed reflection direction, demonstrating effective wavefront manipulation within the 3D geometry. Specifically, a peak gain of 5.59 dBi is achieved at $\theta = 0°$ and $\varphi = 90°$, confirming that the cube assembly does not degrade the beamforming capability of the individual RIS surface. In contrast, when the RIS is switched to the OFF state, the reflected pattern gain is significantly reduced, and the gain drops to −9.96 dBi at $\theta = 0°$. This pronounced contrast between the ON and OFF states verifies the effectiveness of the subarray-level amplitude gating mechanism in the 3D configuration.

Beyond single-surface reflection, the proposed 3D-RIS enables controlled EM signal transfer between neighboring surfaces, which represents a key advantage over conventional planar RIS designs. Fig. 11 illustrates the neighboring-surface transmission scenario, where the incident wave impinges on one face of the cube and is subsequently guided toward an adjacent face through switch-controlled inter-surface connections. In this case, the receiving region on the illuminated surface captures the incident wave and routes it toward the reflection region of the neighboring surface, which then re-radiates the wave toward the desired direction. This mechanism allows the 3D-RIS to redirect signals across different spatial directions without requiring direct illumination of the reflecting surface.

The corresponding simulated radiation characteristics for the neighboring-surface transmission case are shown in Fig. 12. When the RIS is in the ON state, a clear reflected beam is observed on the adjacent surface, with a peak gain of 0.13 dBi at $\theta = 90°$ and $\varphi = 180°$. This result confirms that effective beamforming can be achieved even when the reflection occurs on a surface that is not directly illuminated. In contrast, the OFF state exhibits decrease of the transmitted and reflected fields, with the gain reduced to −16.32 dBi at $\theta = 0°$. The substantial gain difference between the ON and OFF states further demonstrates the robustness of the inter-surface switching mechanism and its ability to dynamically control wave propagation paths within the 3D-RIS structure.

Overall, the results in Fig. 9 through Fig. 12 collectively verify that the proposed 3D-RIS architecture successfully extends conventional planar RIS functionality into the volumetric domain. By combining single-surface beamforming, inter-surface energy transfer, and subarray-level control, the proposed design enables flexible spatial wave manipulation across multiple directions. This 3D capability provides a promising solution for overcoming line-of-sight blockage and coverage limitations in future mm-Wave and sub-THz wireless systems, while maintaining a modular and hardware-efficient implementation.

### C. Control Circuit

The schematic of the control circuit employed in the proposed 3-D RIS is illustrated in Fig. 13. A Skoll Kintex-7 FPGA module is adopted as the digital controller to generate reconfigurable high-level and low-level logic signals corresponding to the ON and OFF states of each RIS subarray. In the implemented system, both the FPGA output logic level and the DC bias supply VCC are set to 3.3 V, enabling a unified low-voltage control architecture with reduced overall power consumption.

Although the FPGA provides sufficient logic-level control, its output current capability is insufficient to directly bias the PIN diodes used in the RIS, particularly when multiple diodes are activated simultaneously. To address this limitation while maintaining low-power operation, a p-channel MOSFET (NCE3401AY) is introduced as a switching element between the 3.3 V DC supply and the PIN-diode bias network. When the FPGA output is pulled low, a sufficiently negative gate-to-source voltage is established, turning the pMOS ON and allowing the required forward bias current to flow through the PIN diodes. Conversely, when the FPGA output is high, the pMOS remains in the OFF state, thereby isolating the PIN-diode network from the bias supply.

A gate resistor of 500 Ω is connected between the FPGA output and the pMOS gate to mitigate switching-induced noise, which is beneficial for stable RIS operation. Series bias resistors are incorporated in each PIN-diode path to regulate the forward bias current and ensure reliable switching behavior. Specifically, a resistance of 330 Ω is used when two PIN diodes are driven in parallel, while a resistance of 40 Ω is adopted for a single PIN-diode path.

### V. FABRICATION AND MEASUREMENT

### A. Fabricated 3D-RIS Prototypes

The fabricated prototypes of the proposed 3D-RIS are presented in Fig. 14, including the single-surface RIS module in Fig. 14(a), the assembled 3D-RIS RF boards in Fig. 14(b), and the complete 3D-RIS system integrated with the control circuitry in Fig. 14(c). The proposed 3D-RIS is fabricated using a modular integration strategy, where each cube surface is first realized as an independent multilayer PCB panel and subsequently mechanically assembled into an orthogonally aligned cubic structure with internal electrical interconnections to the control circuitry. These prototypes are realized to experimentally validate the feasibility of the proposed architecture and to demonstrate the practical implementation of the subarray-based 3D-RIS design discussed in the previous sections.

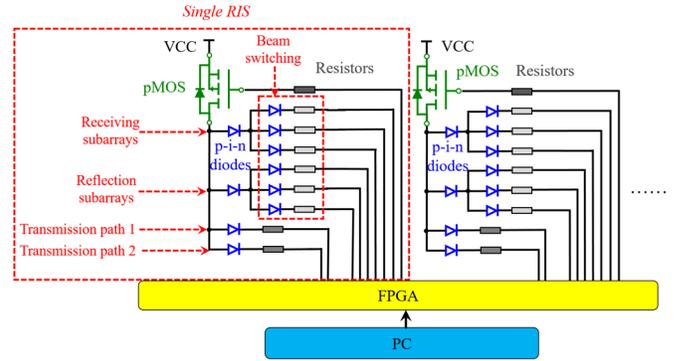

Fig. 13. The control circuit design for the 3D-RIS.

The fabricated single-surface RIS module (shown in Fig. 14(a)), which consists of two separate boards. The first one is the radiating RIS board and the second one is a dedicated ground-plane board, which is mechanically assembled beneath the radiating board. The radiating RIS board is implemented using a multilayer printed circuit board stackup, where a 0.787-mm-thick Rogers 5880 substrate hosts the radiating patch array, and the RF slot layer is implemented on a 0.127-mm-thick Rogers 6002 substrate. These two Rogers boards are bonded using a 0.101-mm-thick RO4450F prepreg. The RF feeding network and DC biasing lines are routed on the bottom surface of the Rogers 6002 board, enabling compact integration of RF routing and bias distribution within the RIS module. For the ground plane board, it provides effective protection for the soldered RF and DC components embedded inside the RIS structure, improving robustness during assembly and handling. More importantly, it serves as an RF shielding layer, suppressing undesired EM leakage and coupling associated with the internal control circuitry, thereby improving the EM isolation and stability of the RIS module. For fabrication quality and long-term reliability, the metallic patterns on the bottom layer of the RIS board are fabricated with solder mask to ensure high-quality component soldering and reliable electrical interconnections. In addition, the top copper layer of the RIS board, as shown in the top view of Fig. 14(a), is treated with a tin-plating process to prevent surface oxidation, which helps preserve conductivity and maintain stable mm-Wave performance.

The individual RF boards (six of them) are assembled into a cube-shaped structure (shown in Fig. 14(b)). DM-SIP-2004 silver electronic paste (DycoTec Materials) has been used to electrically connect neighboring RIS. The mechanical support structures are fabricated using a Bambu Lab H2D 3D printer with PLA filaments, providing sufficient rigidity while maintaining lightweight construction. These printed supports ensure accurate alignment of individual surfaces and stable cube assembly, which is important for preserving the designed inter-surface geometry and ensuring consistent EM behavior across different faces.

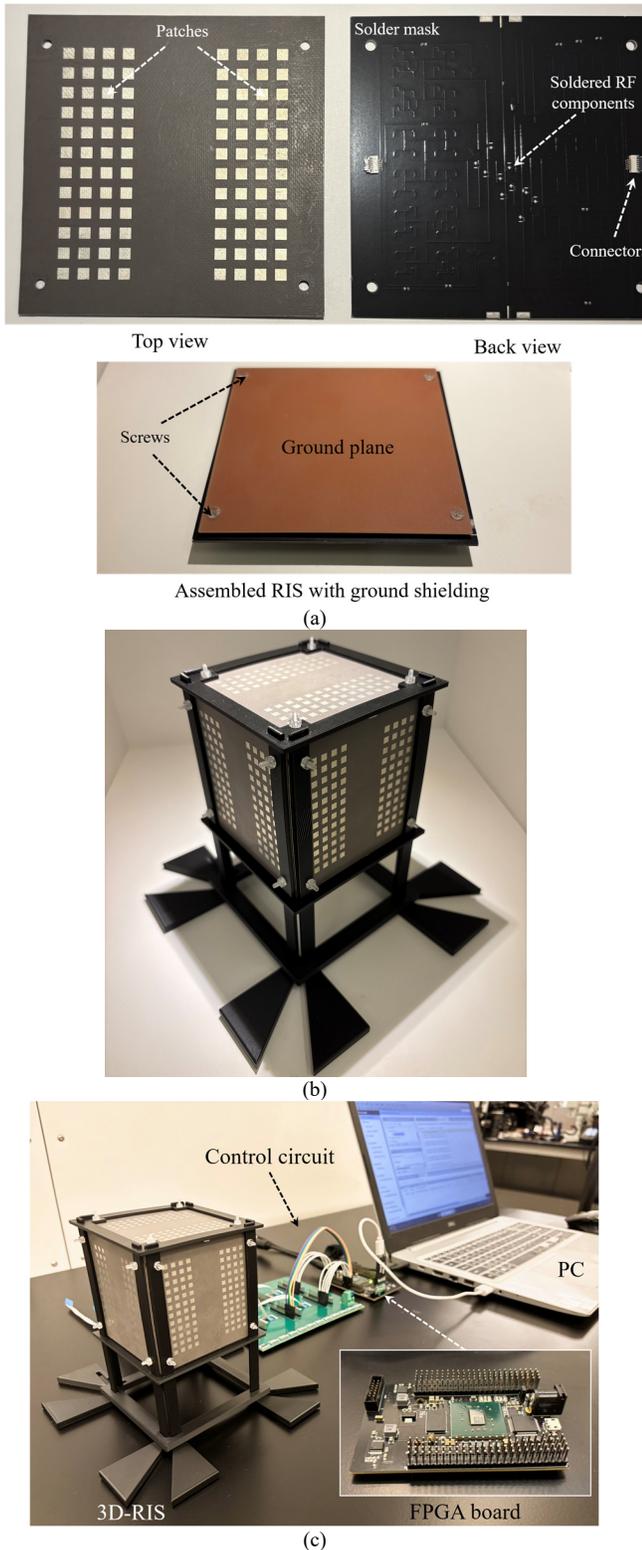

Fig. 14. Fabricated 3D-RIS prototype. (a) Single-surface RIS. (b) 3D-RIS RF boards. (c) The complete 3D-RIS with control circuits.

Fig. 14(c) illustrates the complete 3D-RIS prototype integrated with the control circuits. The RIS are mounted on the printed supports, and flexible flat cables and connectors are employed to interface each RIS surface with the external FPGA-based control hardware. This configuration enables independent and reconfigurable control of the subarray states on each face while maintaining a compact and modular hardware layout. It should be pointed out that, for future versions, the control circuit can be designed to be accommodated inside the 3D-RIS, where the closed ground plane can provide effectivce EM shielding between the external RIS and internal control circuit.

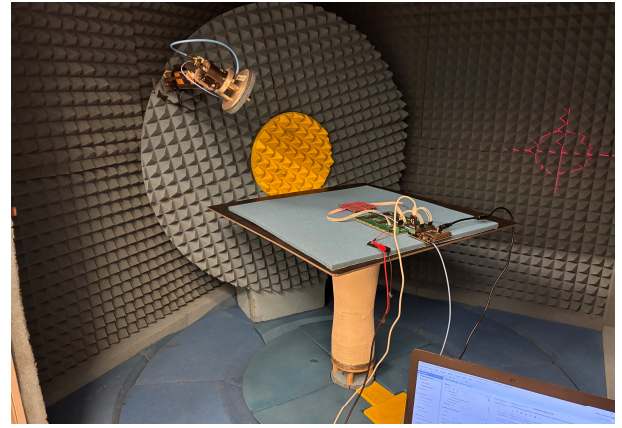

Fig. 15. Measurement setup of a single RIS in an anechoic chamber.

### B. Single surface RIS measurement in the Chamber

To experimentally validate the beamforming and beam-switching performance of the proposed single-surface RIS, far-field measurements are conducted in in an Orbit *u*-Lab mm-Wave anechoic chamber. The measurement setup is illustrated in Fig. 15, where the fabricated single-surface RIS module is mounted on a piece of absorber. The RIS is measured at multiple frequencies spanning from 24 GHz to 30 GHz with a frequency step of 2 GHz to evaluate its broadband performance. The PIN-diode biasing states are controlled by the external FPGA-based control circuit, enabling real-time switching between different subarray activation corresponding to the beamforming configurations discussed in Section III and Section IV-A.

Fig. 16 presents a comprehensive comparison between the simulated and measured beamswitching performance of the single-surface RIS. Figs. 16(a)–(d) show the normalized radiation patterns of the receiving arrays at 24 GHz, 26 GHz, 28 GHz, and 30 GHz, respectively, while Figs. 16(e)–(h) report the corresponding results for the reflection arrays under the same operating frequencies. For each frequency, multiple beam states are generated by selectively activating different subarrays, producing distinct main-beam directions consistent with the designed beamswitching angles (-30°, -15°, 0°, 15°, 30°).

From the measured results, it can be observed that the main-beam directions agree well with the simulated predictions across the entire frequency band. The measured patterns demonstrate discrete beam switching behavior, verifying the effectiveness of the subarray-level amplitude gating mechanism implemented by the PIN diodes. Moreover, the reflection arrays exhibit beam characteristics that follow those of the receiving arrays, which is consistent with the polarization-orthogonal design principle described in Section II. Minor discrepancies between the simulated and measured results are observed in terms of sidelobe levels and beamwidth, which can be mainly attributed to fabrication tolerances and assembling errors. Nevertheless, the overall beam shapes and directions remain in good agreement, confirming the robustness of the beasmswitching capability for the proposed single-surface RIS design.

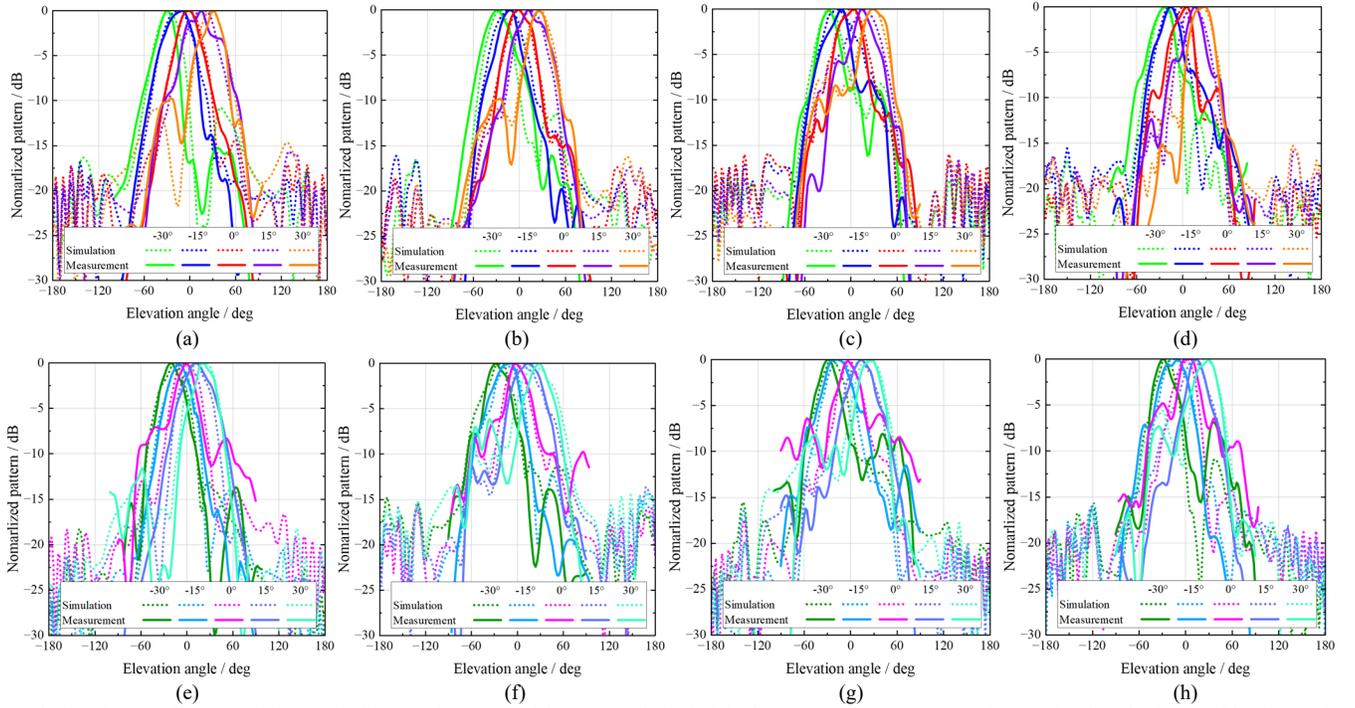

Fig. 16. Simulated and measured beamswitching performance of the designed single RIS for the receiving arrays at (a) 24 GHz, (b) 26 GHz, (c) 28 GHz, and (d) 30 GHz, and for the reflection arrays at (e) 24 GHz, (f) 26 GHz, (g) 28 GHz, and (h) 30 GHz.

*C. 3D-RIS measurement*

The measurement setup for the proposed 3D-RIS with one-sided illumination and reflection is shown in Fig. 17. A PE9851B-20 WR-34 standard gain horn antenna is used as the transmitter to illuminate one surface of the 3D-RIS, while another horn antenna is employed as the receiver to capture the reflected signal along the desired observation direction. In the experiment, both the Tx and Rx horns are positioned at a distance of 12 cm from the illuminated RIS surface, which is around far-field distance for 4×4 subarray, at an incident angle of -30°, forming a one-sided reflection configuration that follows the simulated scenario in Fig. 9. The FPGA-controlled PIN-diode bias network enables real-time switching between the RIS ON and RIS OFF states, so the enhancement introduced by the programmed subarray activation can be directly quantified by comparing the measured $S_{21}$ levels between two horns by a PNA Network Analyzer E8363C.

Fig. 18(a)–(c) present the measured $S_{21}$ parameters at 26 GHz for beam reflection directions of 0°, 15°, and 30°. For the RIS ON state, the measured $S_{21}$ values at the desired reflection directions are −26.2, −26.5, and −25.9 dB, respectively. In contrast, the corresponding $S_{21}$ values for the RIS OFF state are −40.9, −40.1, and −35.6 dB. As a result, the measured gain enhancements introduced by the activated 3D-RIS are 14.7,

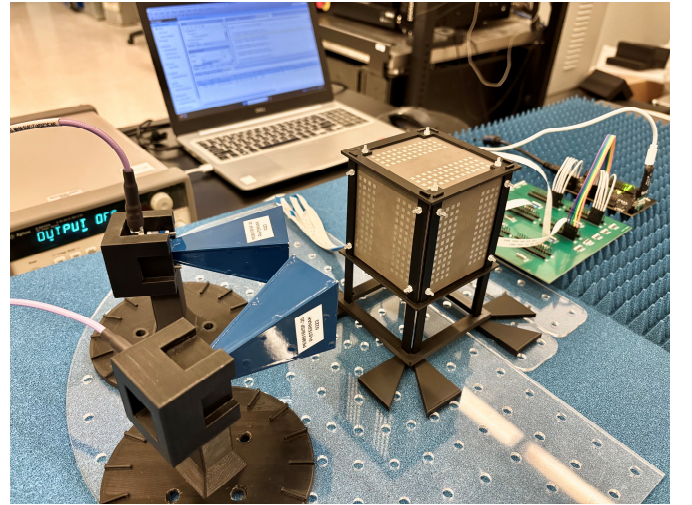

Fig. 17. Measurement setup of the designed 3D-RIS with one-sided illumination and reflection.

13.6, and 9.7 dB for the 0°, 15°, and 30° beam states, respectively. It should be pointed out that the beam switching angles of -15° and -30° cannot be conducted since the Tx horn and the Rx horn collide within such angles. Nevertheless, these results demonstrate the effective beam switching capability of the 3D-RIS with one-sided illumination and reflection under different reflection angles.

To further evaluate the broadband performance of the proposed structure, Fig. 18(d)–(f) show the measured $S_{21}$ responses at 24, 28, and 30 GHz for the 0° beam state. When the RIS is turned ON, the measured $S_{21}$ values are −26.1, −27.9, and −27.5 dB at 24, 28, and 30 GHz, respectively. For comparison, the corresponding RIS OFF values are −39.2, −42.7, and −40.2 dB. Therefore, the measured gain enhancements achieved by the 3D-RIS are 13.1, 14.8, and 12.7 dB across the considered frequency band. Overall, the measured results in Fig. 18 show consistent contrast between the RIS ON and RIS OFF states for both angular beam switching and frequency variation. The measured enhancement levels and their variation trends exhibit with the simulated results presented in Fig. 9 and Fig. 10, validating the effectiveness of the proposed 3D-RIS architecture and its integrated control scheme.

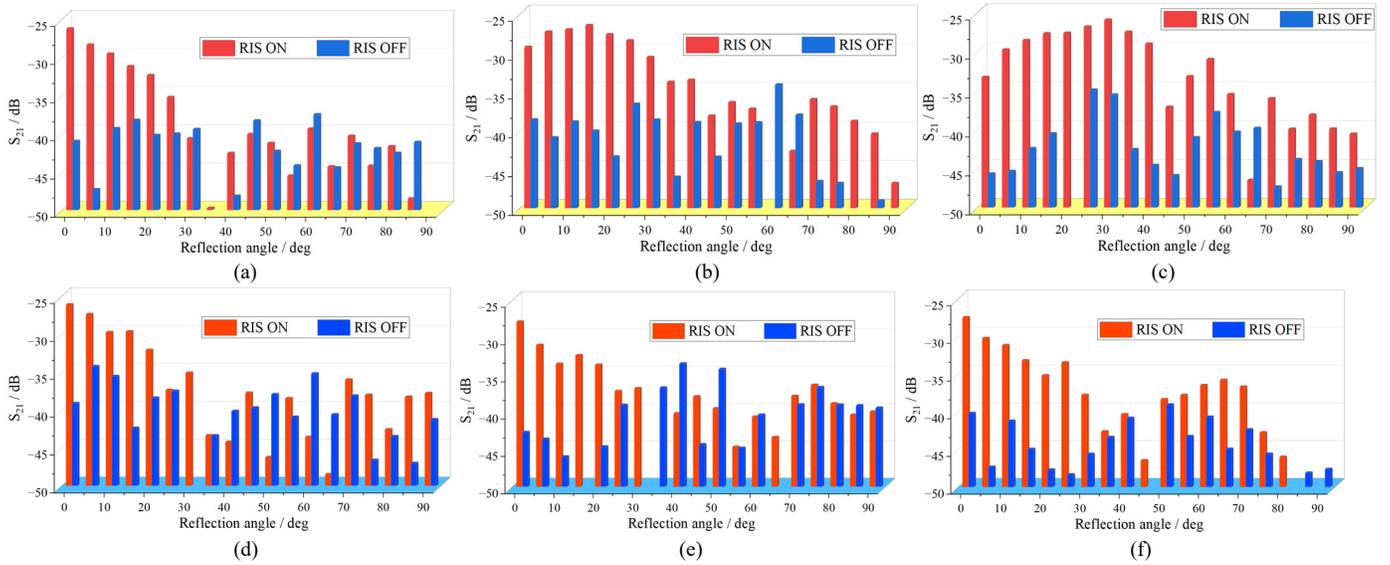

Fig. 18. Measured $S_{21}$ between the Tx and Rx horns for the designed 3D-RIS in ON and OFF states under reflection operation at 26 GHz with beam-switching directions of (a) 0°, (b) 15°, and (c) 30°, and at different frequencies under 0° beamforming of (d) 24 GHz, (e) 28 GHz, and (f) 30 GHz.

Fig. 19 illustrates the measurement setup for the proposed 3D-RIS under transmission opeartion. In this configuration, the incident wave illuminates one face of the cube, while the reflected signal is re-radiated from the adjacent face and received by the Rx horn. The experimental setup follows the simulated neighboring-surface transmission scenario shown in Fig. 11, including the relative orientation of the cube faces, the illumination direction, and the Tx–RIS–Rx geometry.

Fig. 20 reports the measured $S_{21}$ parameters at 26 GHz for beam transmission directions of −30°, −15°, 0°, 15°, and 30°. For the RIS ON state, the measured $S_{21}$ values at the desired transmission directions are −42.4, −41.8, −40.5, −41.9, and −43.1 dB, respectively. In contrast, the corresponding RIS OFF state $S_{21}$ values are −53.8, −53.9, −54.6, −52.6, and −54.7 dB. Accordingly, the measured gain enhancements achieved by activating the neighboring-surface transmission path are 11.4, 12.1, 14.1, 10.7, and 11.6 dB for the five beam states, respectively. Overall, the measured results demonstrate a consistent contrast between the RIS ON and OFF states across various beamswitching transmission angles, confirming the effectiveness of the gain enhancement of the neighboring surfaces of the proposed 3D-RIS. The measured transmission trends and enhancement levels generally match the simulated results in Fig. 12, where a directional pattern is formed on the neighboring surface under the ON state while the transmitted and reflected fields are significantly suppressed under the OFF state. This agreement further validates the beamforming mechanism of the proposed 3D-RIS.

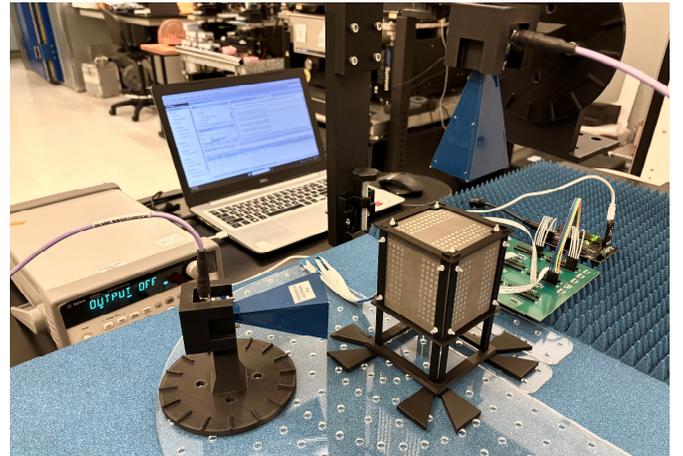

Fig. 19. Measurement setup of the designed 3D-RIS with one-sided illumination and neighboring-side transmission.

### D. Communication Trials

To further validate the proposed 3D-RIS from a communication-system perspective, over-the-air wireless communication trials are conducted using a mm-Wave software-defined radio (SDR) platform combined with external frequency up-conversion and down-conversion modules. The experimental setup is illustrated in Fig. 21(a). A pair of Pluto SDRs is employed, where one operates as the transmitter and the other as the receiver. Both SDRs are configured with a local oscillator (LO) frequency of 2 GHz and generate complex baseband signals with a sampling rate of 512 kS/s. At the transmitter side, the baseband signal generated by the Pluto SDR with 2 GHz and then mixed with a 6 GHz signal generated by an external signal generator through a frequency quadrupling up converter, resulting in a RF carrier centered at 26 GHz. The upconverted 26 GHz signal is subsequently radiated by a standard-gain horn antenna (PE9851B-20) toward the illuminated surface of the proposed 3D-RIS. The receiver adopts an identical architecture in reverse, where the received

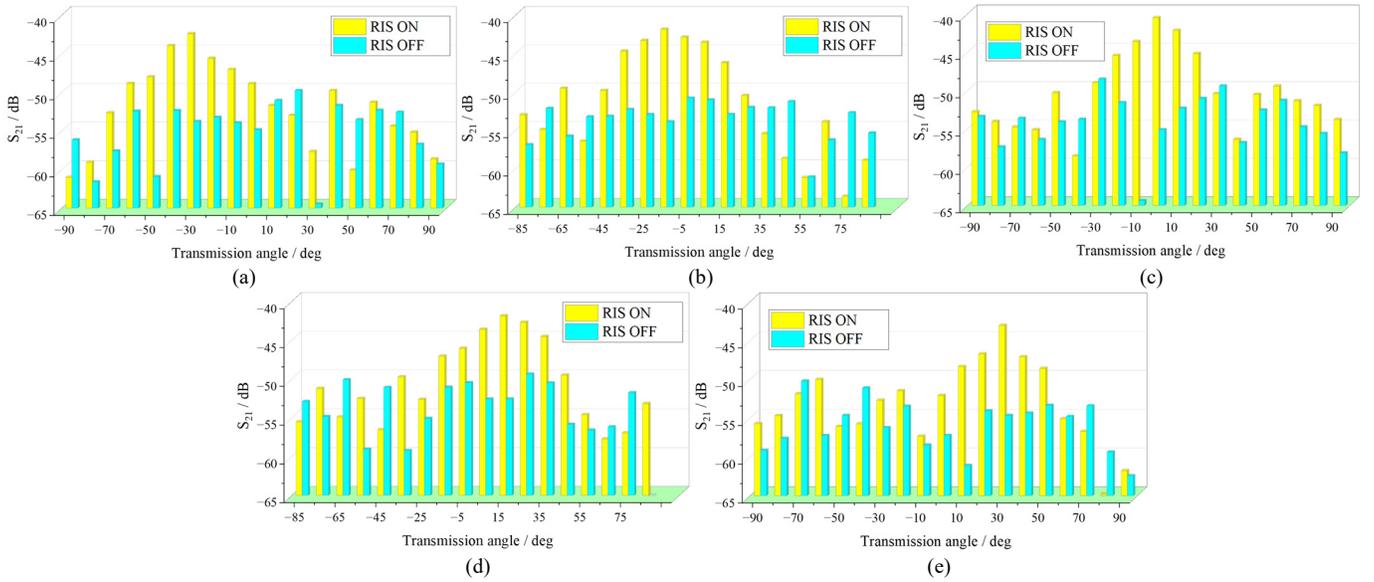

Fig. 20. Measured $S_{21}$ between the Tx and Rx horns for the designed 3D-RIS in ON and OFF states under transmission operation at 26 GHz with beam-switching directions of (a) -30°, (b) -15°, (c) 0°, (d) 15°, and (e) 30°.

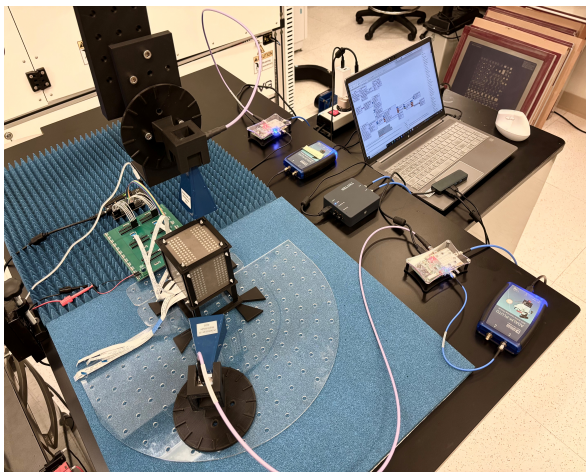
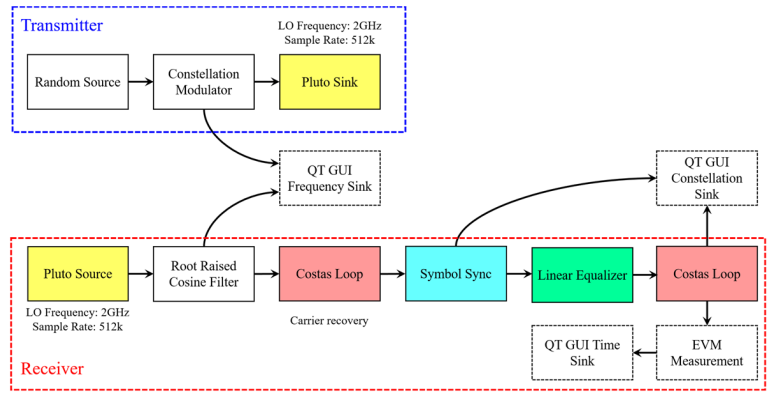

(a) (b)

Fig. 21. Experimental setup for the proposed 3D-RIS-assisted wireless communication system. (a) System-level hardware configuration. (b) Communication software architecture.

26 GHz signal collected by another horn antenna is down converted to an intermediate frequency through the same frequency mixing scheme and finally demodulated by the Pluto SDR. This symmetric transceiver configuration ensures consistent frequency translation and minimizes hardware-induced asymmetry in the communication link.

The communication software architecture implemented in GNU Radio is shown in Fig. 21(b). At the transmitter, a QPSK modulation scheme is adopted, where a random binary source is mapped to complex symbols through a constellation modulator. Root-raised-cosine pulse shaping is applied to control the occupied bandwidth and suppress intersymbol interference before transmission. At the receiver, the incoming signal undergoes matched filtering followed by carrier recovery and symbol synchronization using Costas loop and timing error detector based blocks. An adaptive linear equalizer is further employed to mitigate residual channel impairments introduced by the wireless link and the RIS-assisted propagation path. Finally, constellation demapping and EVM computation are performed to quantitatively evaluate the communication performance. This software architecture allows real-time visualization of the received spectrum, constellation diagram, and EVM, enabling direct comparison between different RIS operating states and propagation scenarios.

The measured communication results are summarized in Fig. 22 for both same-surface reflection and neighboring-surface transmission cases under illumination. When the 3D-RIS is switched to the ON state, a clear improvement in spectral concentration and constellation clustering is observed compared to the OFF state, indicating effective enhancement of the received signal quality (for both reflection and transmission cases). For the reflection scenario, the measured EVM is reduced from average −16.3 dB in the RIS OFF state to average −23.3 dB in the RIS ON state, corresponding to an improvement of 7 dB. This enhancement is accompanied by a significantly tighter QPSK constellation with reduced symbol dispersion and phase noise. For the neighboring-surface transmission case, the proposed 3D-RIS still provides notable communication gains. The measured average EVM is -14.7 dB at RIS OFF while increase to -20.9 dB at ON state, which means a 6.2 dB enhancement when the RIS is activated, and the constellation diagram exhibits well-defined symbol clusters, confirming reliable signal transmission through the inter-surface propagation path. These results are consistent with the radiation measurements reported in the previous sections, and they demonstrate that the proposed 3D-RIS not only enhances EM performance but also translates these gains into tangible improvements at the communication-system level. Overall, the communication trials verify the effectiveness of the proposed 3D-RIS architecture in practical mm-Wave wireless links and highlight its potential for improving link quality in both reflection-dominated and neighboring-surface transmission scenarios.

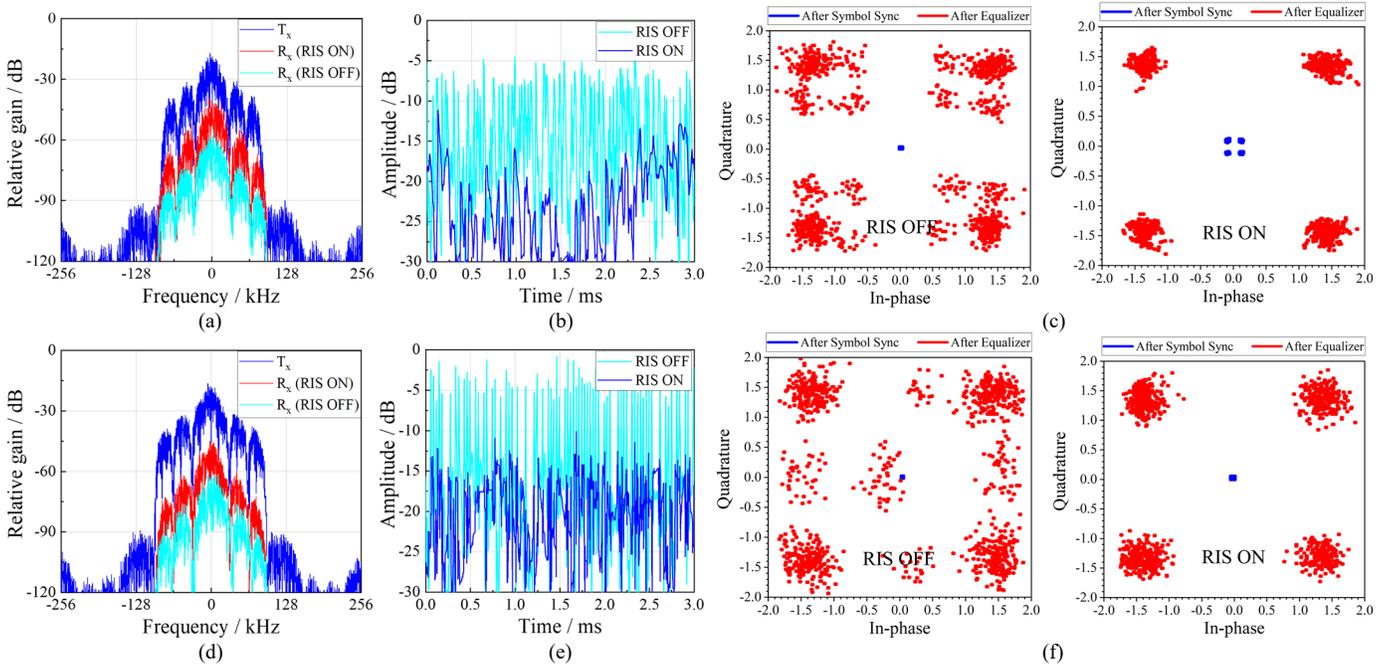

Fig. 22. Measured frequency spectrum, EVM, and constellation for the proposed 3D-RIS-assisted wireless communication system under one-sided illumination. (a)–(c) Same-surface reflection and (d)–(f) neighboring-surface transmission, respectively.

## VI. Conclusion

This work presents a 3D-RIS architecture that extends conventional planar RIS designs into the volumetric domain by enabling both single-surface reflection and controlled neighboring-surface transmission. A unified theoretical framework is developed and implemented through a practical mm-Wave prototype operating over the 24 – 30 GHz band, where subarray-based beamforming with binary amplitude gating and predefined phase offsets achieves flexible beam switching with reduced control complexity. Experimental results demonstrate measured gain enhancements of up to 14.7 dB for one-sided reflection and 14.1 dB for neighboring-surface transmission at 26 GHz, together with consistent broadband performance. Over-the-air communication trials further confirm clear improvements in constellation quality and error vector magnitude, validating that the electromagnetic advantages of the proposed 3D-RIS directly translate into communication-level gains. Future work will focus on extending the proposed theoretical and hardware framework toward more 3D-RIS RIS architectures, including arbitrary polyhedral and non-cubic geometries, as well as exploring advanced inter-surface coupling and joint surface-level synthesis to fully exploit the degrees of freedom enabled by volumetric RIS configurations.